# Do Birds of a Methodological Feather Flock Together?

Carrie E. Fry, PhD & Laura A. Hatfield, PhD

June 19, 2020

Carrie E. Fry, MEd PhD was a graduate student in the Interfaculty Initiative in Health Policy at Harvard University. Laura Hatfield, PhD is an associate professor of Health Care Policy (Biostatistics) in the Department of Health Care Policy at Harvard Medical School.

**Abstract:**

Quasi-experimental methods have proliferated over the last two decades, as researchers develop causal inference tools for settings in which randomization is infeasible. Two popular such methods, difference-in-differences (DID) and comparative interrupted time series (CITS), compare observations before and after an intervention in a treated group to an untreated comparison group observed over the same period. Both methods rely on strong, untestable counterfactual assumptions. Despite their similarities, the methodological literature on CITS lacks the mathematical formality of DID. In this paper, we use the potential outcomes framework to formalize two versions of CITS — a general version described by Bloom (2005) and a linear version often used in health services research. We then compare these to two corresponding DID formulations— one with time fixed effects and one with time fixed effects and group trends. We also re-analyze three previously published studies using these methods. We demonstrate that the most general versions of CITS and DID impute the same counterfactuals and estimate the same treatment effects. The only difference between these two designs is the language used to describe them and their popularity in distinct disciplines. We also show that these designs diverge when one constrains them using linearity (CITS) or parallel trends (DID). We recommend defaulting to the more flexible versions and provide advice to practitioners on choosing between the more constrained versions by considering the data-generating mechanism. We also recommend greater attention to specifying the outcome model and counterfactuals in papers, allowing for transparent evaluation of the plausibility of causal assumptions.

## Section I: Introduction

Methods for causal inference using observational data have proliferated in recent decades, as researchers seek to estimate causal quantities when randomization is infeasible.[1–5] Difference-in-differences (DID) is particularly popular thanks to its apparent conceptual simplicity. Many of the more than 300 papers evaluating the Affordable Care Act's Medicaid expansion use DID.[6] DID compares the change in outcome before and after treatment in a treated group to the change over the same period in a comparison group that does not receive treatment. Comparative interrupted time series (CITS; also known as interrupted time series with a control or controlled interrupted time series) is superficially similar to DID. It also uses treated and comparison groups to compare changes before and after an intervention. The methods are so similar that some have asserted that they are the same.[7,8]

However, proponents of DID and CITS each strongly prefer their respective methods, and these preferences tend to be disciplinary. Clinical epidemiology and education policy researchers gravitate to CITS, while economists and health policy researchers tend to use DID. Causal inference texts in economics may exclude interrupted time series or comparative interrupted time series.[9,10] In contrast, a recent paper about quasi-experimental methods suggested that DID is "weak design" while CITS is a "strong design".[11] In dueling commentaries in the International Journal of Clinical Epidemiology, economists[12] argued that a paper on rigorous CITS was inadequate because it failed to cite the DID literature. They claimed CITS was another name for DID and concluded that this omission misled readers about identification concerns raised in the DID literature. In a rebuttal, the authors of the original paper maintained that CITS is "regarded as a more powerful design than DID."[13]

Both DID and CITS come in many flavors, and, to add to the confusion, education policy and health services researchers often define CITS differently. The education policy literature defines CITS more generally than the health services literature. In this paper, we discuss two common versions of each design. For DID, we consider DID with time fixed effects ("FE DID") and DID with both time fixed effects and group-specific pre-trends ("FE DID with group trends"). For CITS, we consider a generalized CITS ("general CITS") and a fully linear version of CITS ("linear CITS").

We briefly describe these methods here, deferring the details to subsequent sections. Linear CITS fits straight lines through the pre-period outcomes of the treated and comparison groups and assumes that the post-period change in the comparison group's intercept and slope would also occur in the treated group in the absence of treatment. General CITS replaces the intercept and slope change assumption with the assumption that the difference between the observed and extrapolated outcomes of the comparison group at each post-period timepoint would hold in the treated group in the absence of treatment. FE DID uses time fixed effects to capture a common time trend and assumes the pre-period difference between treated and comparison groups would remain the same in the post-period in the absence of treatment. FE DID with group time trends adds a linearly growing difference between the two groups, which it extrapolates from the pre- to post-period (Figure 1). All four target estimands are the average effect of treatment on the treated (ATT).

In some situations, only one method is feasible. FE DID requires only two observation points (one in the pre-period and one in the post-period), while FE DID with group trends and general CITS require at least five observation points (at least four pre and one post), and linear CITS requires at least eight (four pre and four post). DID can estimate a single treatment effect or a time-varying effect at each post-period time point ($ATT_{(t)}$). General CITS also estimates the ATT at each post-period time point, while linear CITS estimates an intercept and slope shift. Many who prefer linear CITS cite the ability to estimate both immediate and sustained effects as an advantage of the method. Linear CITS does easily give the magnitude and direction of the growing treatment effect, while DID and general CITS's time-varying treatment effects are not as amenable to easy summaries.

Critics of DID state say the method is overly simplistic, restrictive, and inflexible because it fails to account for historical trends in the outcome.[11,13,14] In the specification of DID with two time points per group, it is true that this design precludes adjustment for (or examination of) differential outcome trends prior to intervention. However, most DID studies include more than two time points. The counterfactual assumption of DID states that the average change in outcome in the two groups from pre- to post-period would have been the same *if not for treatment*. In practice, most researchers typically impose a more restrictive version of this assumption — the parallel evolution of untreated potential outcomes at all pre- and post-period time points, referred to as the "parallel trends" assumption. However, several alternative model specifications for DID relax this assumption,[15,16] including modeling group-specific trends.[9]

All four designs assume additivity — that one can add the change in the comparison group to the treatment group's pre-period outcomes to construct its counterfactual untreated post-period outcomes. But their linearity assumptions vary. FE DID does not require any parametric model for the outcome. And FE DID with group trends assumes only that the *difference* between the two groups grows linearly. By contrast, both versions of CITS assume that the outcome trend is linear: general CITS assumes linearity in the pre-period trend only, while linear CITS assumes linearity in both the pre- and post-period trends. Critics of CITS argue that it makes unnecessary linearity assumptions. The upside, however, is a way out of DID's parallel pre-trends restriction, because CITS explicitly models differential outcome trends. CITS extrapolates pre-period outcome trends to construct post-period counterfactual outcomes. Deviations from linearity in either period can lead to bias.[17]

In its simplest version, DID can be estimated with the 2x2 table of treated/comparison and pre-/post-period average outcomes. However, most applied DID studies use multiple observations in the pre- and post-periods and time fixed effects to account for overall outcome trends. With continuous outcomes, DID models can be estimated via linear regression or non-parametric approaches, while general and linear CITS require linear regression. Non-linear regression models for other outcome types are substantially more complicated to interpret, and we do not address them in this paper. Table 1 summaries the basics of functional form, extrapolation, and treatment effect estimation.

Despite these critiques and differences, the literature does not offer clear guidance on instances where one design is more appropriate. This paper formalizes DID and CITS to enable a clear comparison. In Section II, we define the untreated potential outcome models of the four designs. In Section III, we provide three empirical examples illustrating the consequences of choosing CITS versus DID. As we demonstrate in Section II and III, the general forms of CITS and DID produce the same counterfactuals and the same treatment effect estimates. The only differences are language and disciplinary culture. In Section IV, we conclude with guidance on how to use each study design and improve reporting of analyses that use these models. We recommend that researchers use the more flexible version of these two designs, unless there is a good reason to choose one of the more constrained versions, and always state their causal and statistical assumptions clearly.

**Section II: Comparison of Study Designs' Potential Outcomes**

We begin by defining potential outcomes and constructing counterfactual outcomes in each design. Suppose the true data-generating model is $y \sim (t/2)^2$ in the comparison group and $y \sim (t/3)^2$ in the treated group. The overall trends in both groups are non-linear and the differential growth between the two groups is also non-linear. The data-generating model does not change from the pre-period to the post-period, and the intervention has no effect on outcomes. We show how different models would construct the treated group's untreated potential outcomes in the post-period (i.e., the counterfactual outcomes). In Figure 1, we plot true outcomes (dots) and model extrapolations (the lines) for the comparison group (blue) and treated group (purple).

Let $Y^0$ be the untreated potential outcome, $t$ be time (with $t < t_0$ indicating the pre-period and $t \geq t_0$ the post-period), $\beta$ be the parameters governing pre-period outcomes, and $\check{\beta}$ be the parameters governing post-period outcomes. We use superscripts to indicate group— 0 for the comparison group and 1 for the treated group. In each study design, the target estimand is the average effect of treatment on the treated ($ATT$), which is defined as the difference between the observed (treated) and counterfactual (untreated) outcomes in the treatment group during the post-period.

**General formulation of CITS**

The following version of CITS is found in Bloom and Riccio's 2005 analysis of the Jobs-Plus program:[18] separate lines are fit through the pre-period outcomes of the comparison and treated groups and then extrapolated into the post-period for both groups. The counterfactual is constructed by first measuring the distance from the extrapolated line to the observed post-period outcomes in the comparison group at each post-period time point, $t \geq t_0$. Then, these distances are added to the extrapolated line for the comparison group at each time point. The comparison group's untreated potential outcomes are

$$E[Y_t^0] = \beta_0^0 + \beta_1^0 t + \sum_{k=t_0}^{T} \breve{\beta}_k^0,$$

and the treated group's untreated potential outcomes are

$$E[Y_t^0] = \beta_0^0 + \beta_0^0 t + \sum_{k=t_0}^{T} \breve{\beta}_k^0 + \beta_0^1 + \beta_1^1 t.$$

The parameters $\beta_0^1$ and $\beta_1^1$ are the differential intercept and slope of the treated group relative to the comparison group. In this version of CITS, we construct a counterfactual outcome at each post-period time $t \geq t_0$ using the $\breve{\beta}_k^0$ parameters. In Panel A of Figure 1, we see that the linear model of CITS is a reasonable fit to the pre-period data of both groups. Although this model has additional flexibility in the post-period to capture the non-linear trend in the comparison group, it does not capture the (non-linearly) growing gap between treated and comparison groups because of the underlying linearity extrapolation.

**Linear Formulation of CITS**

The fully linear version of CITS also fits lines through each group's pre-period outcomes and extrapolates those lines into the post-period. Instead of measuring the distance from each point in the post-period to the extrapolated line, linear CITS fits another line through the comparison group's post-period outcomes. The counterfactual is constructed by adding the comparison group's pre-to-post differences in intercept and slope to the linear extrapolation of the treated group. The comparison group's untreated potential outcomes are

$$E[Y_t^0] = \beta_0^0 + \beta_1^0 t + (\breve{\beta}_0^0 + \breve{\beta}_1^0 t) I_{\{t \geq t_0\}}$$

and the treated group's untreated potential outcomes are

$$E[Y_t^0] = \beta_0^0 + \beta_1^0 t + (\breve{\beta}_0^0 + \breve{\beta}_0^0 t) I_{\{t \geq t_0\}} + \beta_0^1 + \beta_1^1 t.$$

Here, $\beta_0^0$ and $\beta_1^0$ are the pre-period intercept and slope for the comparison group, $\breve{\beta}_0^0$ and $\breve{\beta}_1^0$ are the change in intercept and slope for the comparison group in the post period, and $\beta_0^1$ and $\beta_1^1$ are the differential intercept and slope in the treated group.

Because the generated outcomes are reasonably linear *within* each group during the pre-period and because the linear version of CITS estimates a different line in each group and study period, the linear version of CITS fits the pre-period outcomes fairly well (Figure 1, Panel B). However, the non-linearity is more apparent in the post-period, as is the non-linearly growing difference between the two groups.

**DID with time fixed effects**

DID with time fixed effects (FE DID) assumes a constant difference between the treated and comparison groups. FE DID measures the average level change from the pre-period to the post-period in the comparison group. To construct the counterfactual, FE DID assumes that the same change would have been seen in the treated group *absent treatment*.

In the formulation of the potential outcomes for FE DID, we use $\gamma_t$ to represent the time fixed effects, which are estimated throughout the pre- and post-period. The untreated potential outcomes in the comparison group are

$$E[Y_t^0] = \beta_0^0 + \breve{\beta}_0^0 + \gamma_t,$$

and the untreated potential outcomes in the treated group are

$$E[Y_t^0] = \beta_0^0 + \breve{\beta}_0^0 I_{\{t \geq t_0\}} + \beta_0^1 + \gamma_t.$$

In this model, $\beta_0^0$ and $\breve{\beta}_0^0$ are the average level of the comparison group in the pre- and post-period, respectively and $\beta_0^1$ is the differential level of the treated group in the pre-period. In Figure 1, Panel C, we see the "parallel pre-trends" or constant difference assumption of DID. Because this, the FE DID fails to capture the growing difference between the two groups.

**DID with time fixed effects and group-specific trends**

Unlike FE DID, FE DID with group-specific trends allows the treated group to have a (linearly) diverging trend relative to the comparison group in the pre-period, which is then extrapolated into the post-period. The linear differential time trend is *net of* the average time trend captured by the time fixed effects. Like FE DID, the analysis assumes that the average level change in the comparison group would occur in the treated group *absent treatment*. But unlike FE DID, the counterfactual is constructed by first extrapolating the pre-period differential trends. The untreated potential outcomes in the comparison group are

$$E[Y_t^0] = \beta_0^0 + \check{\beta}_0^0 I_{\{t \geq t_0\}} + \gamma_t,$$

and the untreated potential outcomes in the treated group are

$$E[Y_t^0] = \beta_0^0 + \check{\beta}_0^0 I_{\{t \geq t_0\}} + \beta_0^1 + \beta_1^1 t + \gamma_t.$$

Like FE DID, this captures a level difference between treated and comparison in the pre-period ($\beta_0^1$) and an average change in the comparison group in the post-period ($\check{\beta}_0^0$). Unlike FE DID, this adds a differential linear trend in the pre-period ($\beta_1^1$). Here, there is only one trend estimate because of the time fixed effects.

In our simulated data, the true data-generating model includes a differential trend in treated and comparison groups, so accounting for this improves the model fit in the pre-period (Figure 1, Panel D). In addition, this model's counterfactual includes the growing differential between the two groups in a way that the FE DID model cannot.

This counterfactual construction is identical to that of general CITS because the fixed effects are not being extrapolated into the post-period. The extrapolation occurring in FE DID with group trends is a differential linear time trend, which is the same extrapolation in general CITS. The difference is the assumption made about the pre-period outcome. In general CITS, the outcome's trend is assumed to be linear, but FE DID with group trends assumes the difference between the groups is growing linearly.

## Section III: Empirical Examples

In the following, we re-reanalyze data from three papers, compare our findings to the original results, explore differences in treatment effects across study designs, and discuss the how researchers might choose between DID and CITS. We fit four models to each dataset 1) FE DID, 2) FE DID with group trends, 3) general CITS, and 4) linear CITS. In both DID formulations, we estimate time-varying treatment effects. In addition to plotting model fits and treatment effects, we make event study plots, which show the adjusted differential change in outcomes between treated and comparison groups in each time period relative to a single pre-period reference time. These provide a useful visual of how outcomes differentially evolve in the two groups during both pre- and post-periods but are not appropriate for formal inference.

**Medicaid expansion's spillover to the criminal justice system**

Fry, McGuire, and Frank[19] conduct three case studies to estimate the impact of the ACA's Medicaid expansion on rates of return to county jails. The rationale for linking medical insurance coverage to jail involvement is the confluence of substance use disorders, serious mental health diagnoses, and uninsurance among people in jails. Each case study compares the change in outcomes for a county in a state that expanded Medicaid to the change in outcomes for a county in a state that did not expand Medicaid. For brevity, we re-analyze only one of the three case studies and only one outcome, the probability of re-arrest.

The authors of the original paper briefly discuss the rationale for choosing a CITS design, "Given the drivers of the outcome in this study (e.g., policing practices, criminal justice practices, and access to behavioral health services) and how they may vary between the counties, assuming linearity in the evolution of the outcomes in each of the two groups seems more reasonable than assuming that they evolve in the same average way over time" (page 16). The authors also provided qualitative information of differential demographics, access to behavioral health resources, policing practices, and coordination between the behavioral health and criminal justice systems in the study counties. Thus, assuming outcomes evolve in the same way in treatment and comparison groups was deemed unreasonable.

The difference in the pre-period trends is large and approximately linear (Figure 2, Panel A). The diverging pre-period outcomes suggest that FE DID is not an appropriate study design but accounting for the pre-period divergence via CITS or FE DID with group trends may be more appropriate. After Medicaid expansion, there is an immediate decrease and a flattening of the differential trend, suggesting that CITS will estimate negative differential changes in both intercept and slope.

Indeed, using linear CITS, we estimate that Medicaid expansion reduces the probability of any re-arrest by 2.00 (95% CI: 1.62, 2.34) percentage points in the first month after expansion (i.e., the differential intercept change), and the decline grows by 0.07 (95% CI: 0.06, 0.08) percentage points each month (i.e., the differential slope change). By contrast, using FE DID, which does not account for the diverging pre-trends, the estimated effect at the midpoint of the post-period is 2.32 percentage points (or 84%) smaller than the other estimates, and the treatment effect grows in a positive direction.

Diverging pre-period trends is one piece; the other is the linearity of the differential evolution in the post-period. As expected, FE DID with group trends and general CITS estimates are identical (Figure 2, Panel B). They also closely track the linear CITS estimates, which is also expected, given the approximately linear evolution of the differential changes in the post-period (Figure 2, Panel A).

Drivers of recidivism, such as policing and behavioral health treatment availability, differ across counties and time. Thus, FE DID's assumption of a (counterfactual) constant difference in expansion and non-expansion counties is not reasonable. By contrast, the differential outcome trends are approximately linear in both pre- and post-periods, so linear CITS, general CITS, and FE DID with group trends all produce similar estimates. The crucial piece in this example is accounting for diverging pre-period trends (with linear CITS, general CITS, or FE DID with group trends). The data that support the findings of this reanalysis are available on request from the corresponding author.

**Medicaid expansion and naloxone prescriptions**

Frank and Fry[20] compare the change total in naloxone prescriptions before and after Medicaid expansion in expansion and non-expansion states. The rationale for linking insurance coverage to naloxone is that increased financial access will increase treatment with naloxone. While the authors do not provide any explicit rationale for choosing DID, they write, "the number of naloxone prescriptions paid for by Medicaid was essentially identical in expansion states compared to non-expansion states" and compare the unadjusted outcome trends for both groups.

The event study plot in Figure 3 (Panel A) shows that the differential trend hovers very close to zero in the pre-period, which supports FE DID's assumption of no differential trends. In the post-period, the outcomes diverge non-linearly, which suggests that a linear CITS model will not capture this dynamic. General CITS and FE DID with group trends both flexibly model the post-period, so we expect that they will produce similar estimates to the FE DID model, given the lack of differential pre-period trends.

Indeed, the linear CITS model estimates a differential *decrease* of 66.8 (95% CI: 7.6, 125.9) naloxone prescriptions in the quarter after expansion, with a growing increase of 25.4 (95% CI: 7.1, 43.7) prescriptions in each quarter thereafter. At the beginning and end of the post-period, the linear CITS estimates differ substantially from the other three models (Figure 3, Panel B). Again, general CITS and FE DID with group trends produce identical estimates (Figure 3, Panel B), and because of the lack of differential pre-trends, the FE DID estimates are also very similar.

Drivers of naloxone prescriptions include the prevalence of opioid use disorder and laws affecting access, such as the ability of doctors to prescribe naloxone to a friend or family member. Although these vary across both states and time, and anecdotal evidence suggests some states expanded Medicaid partly in response to growing opioid use disorder, the event study plot shows no differential evolution of naloxone prescriptions in the pre-period. However, that plot does show non-linear differential trends in the post-period. The crucial piece in this example is flexibly modeling the post-period differential trends (with general CITS, FE DID, or FE DID with group trends). Data and code for this re-analysis are archived here: https://doi.org/10.7910/DVN/47TMEO.

**Reformulation of OxyContin and the incidence of Hepatitis C**

Powell, Alpert, and Pacula[22] explore the relationship between the 2009 reformulation of OxyContin to an abuse-deterrent form and changes in the incidence of acute Hepatitis C (HCV) infections. The rationale for linking OxyContin reformulation to HCV is that people will switch from OxyContin to illicit opioids, including intravenous heroin, and be exposed to bloodborne illnesses like HCV. The reformulation of OxyContin was a national policy implemented in all U.S. states, so there is no unexposed comparison group. Instead, the authors used a continuous measure of each state's rate of OxyContin abuse or misuse in the pre-period, reasoning that this is a proxy for the "strength" of the reformulation's impact. Using a continuous exposure variable assumes that the relationship between initial OxyContin abuse or misuse and the treatment impact is linear. In addition, the original study drew inferential conclusions from the event study, but the counterfactual assumptions of this approach were not formalized.

The authors write, "The testable assumption is that OxyContin misuse rates were not predictive of hepatitis C infection trends before the reformulation. Studying the effect of a policy exposure both before and after the intervention in an event study is recommended when using difference-in-differences designs to study health policy." (page 289) While counterfactual assumptions are *not* testable, this language suggests the DID parallel pre-trends assumption and the procedures often used to "test" this assumption. Powell, et al. conclude that the reformulation of OxyContin did not change the HCV cases in the year after formulation but did in each subsequent year (see Exhibit 4 of the original paper).[22]

Panel A of Figure 4 suggests modest differential pre-period trends and some non-linearities in the post-period differential evolution. Together, these suggest that both accounting for pre-period trends and flexibly modeling the post-period differential evolution will be important. Thus, we expect all four models will estimate similar treatment effects.

Using FE DID, we estimate an increase in the rate of HCV infections in years 2 through 5 after reformulation (Figure 4, Panel B). When we account for diverging pre-period trends with general CITS and FE DID with group trends, our estimates get slightly smaller in years 4 and 5 after reformulation. The decrease in estimates when we model group-specific pre-period trends demonstrates the impact of differential trends, even when they are not statistically different from zero.

Linear CITS also accounts for the difference in pre-period trends and thus also results in smaller effect estimates than the FE DID. Although the linear CITS estimates look fairly similar to those produced by the other three study designs, modeling the post-period trend as linear misses the slight increase in years 2 and 3 after reformulation that the more flexible modeling of FE DID and general CITS. The result is that neither the intercept nor slope estimates in linear CITS are statistically significant.

Prior to the reformulation of OxyContin, there were small differential trends in HCV rates, perhaps driven by differential rates of intravenous drug use, risky sexual behavior, and access to services for the detection and treatment of HCV, which all vary across states and time. It is unclear whether acute HCV incidence would have evolved similarly in "high" and "low" exposure states in the absence of reformulation. In our re-analyses, general and linear CITS and FE DID with group trends all accounted for the divergent pre-period trends, but linear CITS missed dynamics in the treatment effect. Accounting for pre-period differential trends and flexibly modeling the treatment effect in the post-period were both important in this example, but the impacts were subtle compared to Fry et al. (2020).[19]

## Section IV: Conclusion

Through mathematical formalization of the models and careful examination of the counterfactual assumptions, we have highlighted the differences and similarities among four models: general CITS frequently used in education policy research, linear CITS used most often in health services research, DID with time fixed effects, and DID with time fixed effects and group-specific trends. The counterfactual is constructed similarly in all four designs — by assuming that the change in the comparison group can stand in for the change in the treated group absent treatment.

In their most general forms, CITS (i.e., general CITS) and DID (i.e., FE DID with group trends) produce the same counterfactuals and estimate the same treatment effects. The only difference between these two designs is disciplinary preferences for the language to describe them. Because of their flexibility to model differential pre-period trends and non-linear evolution of the treatment effect, we suggest that researchers consider general CITS or FE DID with group trends first.

When we lean into each design's respective constraints (linearity for CITS and a constant difference for DID), the counterfactuals begin to differ. The choice between these two less-flexible models should rely on content expertise about the data-generating model. If one believes the evolution of the outcome and its confounders is linear over the entire study period, then linear CITS may be preferable. If one believes that the differential evolution in the two groups' untreated potential outcomes is zero, FE DID may be preferable. As recommended by Bilinksi and Hatfield (2020)[23], one way to assess the plausibility of the "parallel pre-trends" assumption of DID is to compare the estimates of FE DID and FE DID with group trends.

While we believe model choice should rely on careful consideration of the counterfactual assumption and the evolution of the outcome, visual devices like adjusted event study plots can be helpful as well. Evidence of adjusted, differential pre-period trends in an event study plot may suggest that a DID may not be appropriate, as seen in the re-analysis of Fry, et al.[19] and Powell, et al.[22]. However, where non-linear outcome trends are present, imposing the linearity assumption of CITS may not fully capture the treatment effect over time as seen in the re-analysis of Frank & Fry[20] and Powell, et al.[22]

In sum, researchers considering a two-group, pre-post study design should opt for a flexible model, whether they call it general CITS or FE DID with group trends, unless there is a strong reason for choosing a more constrained version. In addition, researchers should provide detailed model specifications and counterfactual assumptions in their empirical work. This allows for a transparent evaluation of the plausibility of these assumptions and clear interpretation of the resulting causal effect estimates.

**Acknowledgements:**

Funding for this work has been provided by the Laura and John Arnold Foundation. Carrie Fry is also supported by NIMH training grant T32MH019733. This work is the sole opinion of the authors and does not represent the views of the NIH or the Arnold Foundation. We would like to thank Sherri Rose, Bret Zeldow, Alyssa Bilinski, Alex McDowell, and Luke Miratrix for their thoughtful comments on earlier versions of this work and David Powell and Rosalie Pacula for providing data and code for one of the re-analyses.

**References**

1. LaLonde RJ. Evaluating the Econometric Evaluations of Training Programs with Experimental Data. *The American Economic Review*. 1986;76(4):604-620.

2. Cook TD, Shadish WR, Wong VC. Three conditions under which experiments and observational studies produce comparable causal estimates: New findings from within-study comparisons. *J Pol Anal Manage*. 2008;27(4):724-750. doi:10.1002/pam.20375

3. St.Clair T, Cook TD, Hallberg K. Examining the Internal Validity and Statistical Precision of the Comparative Interrupted Time Series Design by Comparison With a Randomized Experiment. *American Journal of Evaluation*. 2014;35(3):311-327. doi:10.1177/1098214014527337

4. Fretheim A, Zhang F, Ross-Degnan D, et al. A reanalysis of cluster randomized trials showed interrupted time-series studies were valuable in health system evaluation. *Journal of Clinical Epidemiology*. 2015;68(3):324-333. doi:10.1016/j.jclinepi.2014.10.003

5. Ferraro PJ, Miranda JJ. Panel Data Designs and Estimators as Substitutes for Randomized Controlled Trials in the Evaluation of Public Programs. *Journal of the Association of Environmental and Resource Economists*. 2017;4(1):281-317. doi:10.1086/689868

6. Antonisse L, Garfield R, Rudowitz R, Guth M. *The Effects of Medicaid Expansion under the ACA: Updated Findings from a Literature Review*. Kaiser Family Foundation; 2019. https://www.kff.org/medicaid/issue-brief/the-effects-of-medicaid-expansion-under-the-aca-updated-findings-from-a-literature-review-august-2019/

7. Wing C, Simon K, Bello-Gomez RA. Designing difference in difference studies: Best practices for public health policy research. *Annual Review of Public Health*. 2018;39(1):453-469. doi:10.1146/annurev-publhealth-040617-013507

8. Craig P, Katikireddi SV, Leyland A, Popham F. Natural Experiments: An Overview of Methods, Approaches, and Contributions to Public Health Intervention Research. *Annu Rev Public Health*. 2017;38(1):39-56. doi:10.1146/annurev-publhealth-031816-044327

9. Angrist JD, Pischke J-S. *Mostly Harmless Econometrics: An Empiricist's Compansion*. Princeton University Press; 2009.

10. Abadie A, Cattaneo MD. Econometric Methods for Program Evaluation. *Annu Rev Econ*. 2018;10(1):465-503. doi:10.1146/annurev-economics-080217-053402

11. Rockers PC, Røttingen J-A, Shemilt I, Tugwell P, Bärnighausen T. Inclusion of quasi-experimental studies in systematic reviews of health systems research. *Health Policy*. 2015;119(4):511-521. doi:10.1016/j.healthpol.2014.10.006

12. Benmarhnia T, Rudolph KE. A rose by any other name still needs to be identified (with plausible assumptions). *International Journal of Epidemiology*. 2019;48(6):2061-2062. doi:10.1093/ije/dyz049

13. Lopez Bernal J, Cummins S, Gasparrini A. Difference in difference, controlled interrupted time series and synthetic controls. *International Journal of Epidemiology*. 2019;48(6):2062-2063. doi:10.1093/ije/dyz050

14. Soumerai SB, Starr D, Majumdar SR. How Do You Know Which Health Care Effectiveness Research You Can Trust? A Guide to Study Design for the Perplexed. *Prev Chronic Dis*. 2015;12:E101. doi:10.5888/pcd12.150187

15. Mora R, Reggio I. Alternative diff-in-diffs estimators with several pretreatment periods. *Econometric Reviews*. 2019;38(5):465-486. doi:10.1080/07474938.2017.1348683

16. Athey S, Imbens GW. Identification and Inference in Nonlinear Difference-in-Differences Models. *Econometrica*. 2006;74(2):431-497. doi:10.1111/j.1468-0262.2006.00668.x

17. Baicker K, Svoronos T. Testing the Validity of the Single Interrupted Time Series Design. *NBER Working Paper No 26080*. Published online July 2019. https://www.nber.org/papers/w26080.pdf

18. Bloom HS, Riccio JA. Using Place-Based Random Assignment and Comparative Interrupted Time-Series Analysis to Evaluate the Jobs-Plus Employment Program for Public Housing Residents. *The ANNALS of the American Academy of Political and Social Science*. 2005;599(1):19-51.

19. Fry CE, McGuire T, Frank RG. Medicaid Expansion's Spillover to the Criminal Justice System: Evidence from Six Urban Counties. *Working Paper*. Published online 2020.

20. Frank RG, Fry CE. The impact of expanded Medicaid eligibility on access to naloxone. *Addiction*. 2019;114(9):1567-1574. doi:10.1111/add.14634

21. Frank RG, Fry CE; 2020; Naloxone and Medicaid expansion data set; GitHub; LINK

22. Powell D, Alpert A, Pacula RL. A Transitioning Epidemic: How The Opioid Crisis Is Driving The Rise In Hepatitis C. *Health Affairs*. 2019;38(2):287-294. doi:10.1377/hlthaff.2018.05232

23. Bilinski A, Hatfield LA. Seeking evidence of absence: Reconsidering tests of model assumptions. *arXiv:180503273 [stat]*. Published online May 8, 2018. Accessed July 23, 2018. http://arxiv.org/abs/1805.03273

**Table 1.** Comparison of Features of CITS and DID

| Design | # Time Points | Functional Form | Extrapolation | Linearity Assumption | Treatment effect estimation |
|---|---|---|---|---|---|
| General CITS | ≥ 5 | Linear, additive | Linear | Pre-period trend | $ATT_{(t)}$; Linear regression |
| Linear CITS | ≥ 8 | Linear, additive | Linear | Pre & post-period trend | Intercept, slope; Linear regression |
| FE DID w/ group trends | ≥ 5 | Linear, additive | Linear[2] | Differential growth between groups | $ATT_{(t)}$; Linear regression/non-parametric |
| FE DID | ≥ 2 | Additive | Zero difference | None | $ATT_{(t)}$; Linear regression/non-parametric |

[1] Can be used to construct an $ATT_{(t)}$
[2] Linearity is assumed for the evolution of the difference between the treated and control, rather than for the outcome trend itself.

**Figure 1.** Comparison of counterfactual scenarios in non-linear models

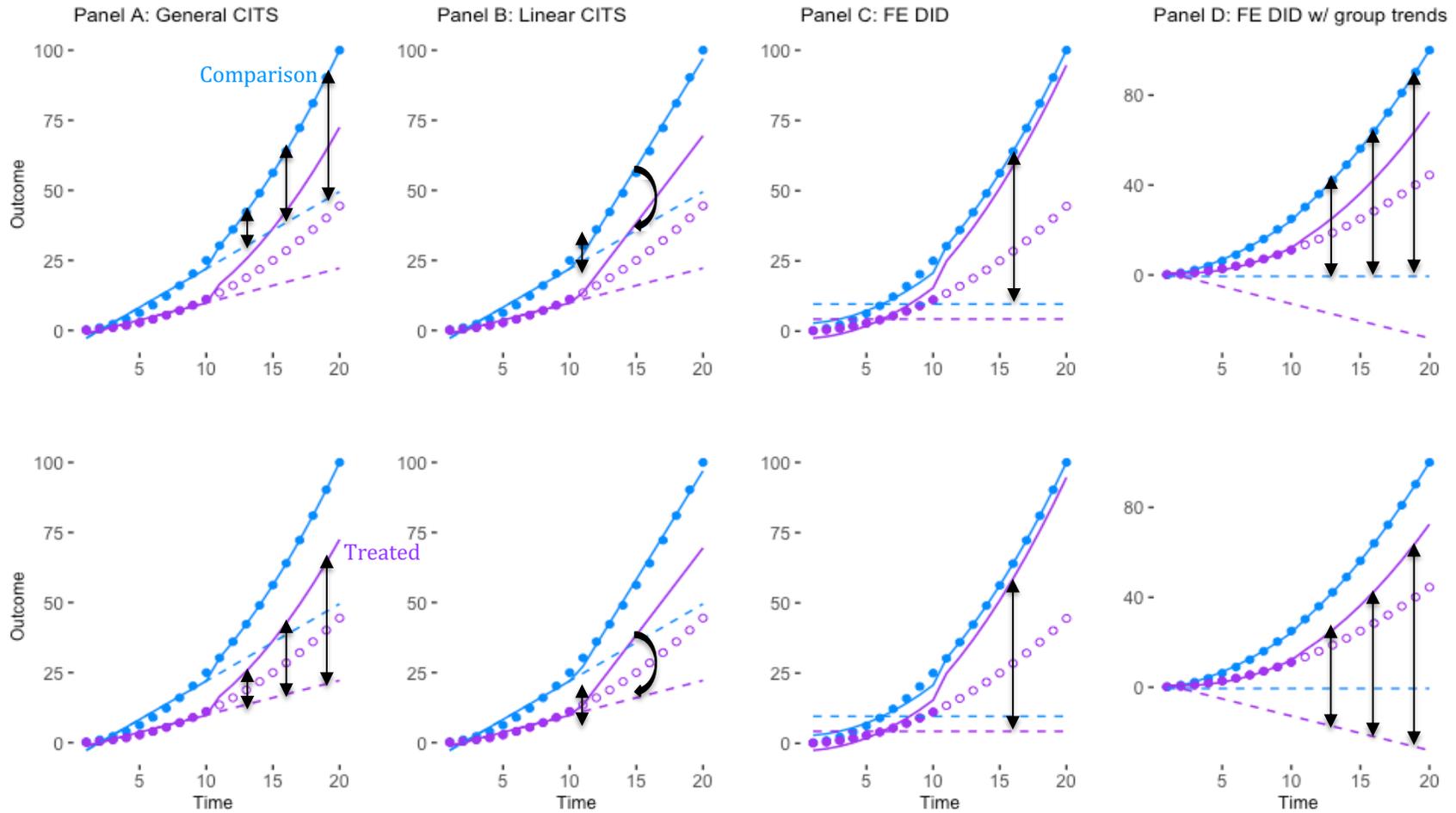

NOTES: The untreated group's true outcomes (blue dots) are generated from the model, $y \sim \left(\frac{t}{2}\right)^2$, and the treated group's true outcomes in the pre-period (purple dots) are generated from the model, $y \sim \left(\frac{t}{3}\right)^2$. The treated group's true, unobservable post-period untreated outcomes are the empty dots. Lines are the modeled outcomes using linear regression for each study design.

**Figure 2.** Comparison of Estimates Across Study Designs - Fry et al, 2020

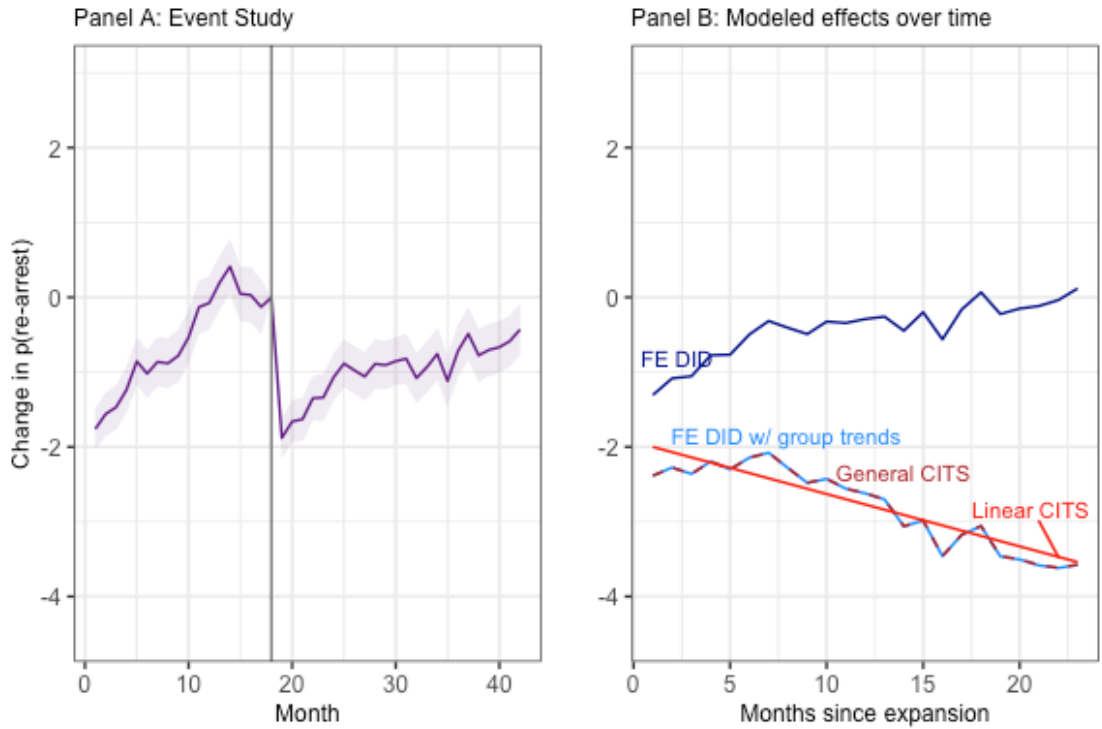

SOURCES/NOTES: **Sources** Authors' re-analysis of booking and release data from Fry et al., 2020. **Notes** Panel A is an event study plot, where the adjusted differential in outcome between the treated group and comparison group relative to the time of intervention is estimated for each time period before and after Medicaid expansion. Panel B provides time-varying estimates for each month after Medicaid for FE DID, FE DID with group trends, general CITS, and linear CITS. Covariate adjustment is the same for each model presented in both Panels A and B and is exactly the same as the covariate adjustment used in Fry et al., 2020.

**Figure 3.** Comparison of Estimates Across Study Designs - Frank & Fry, 2019

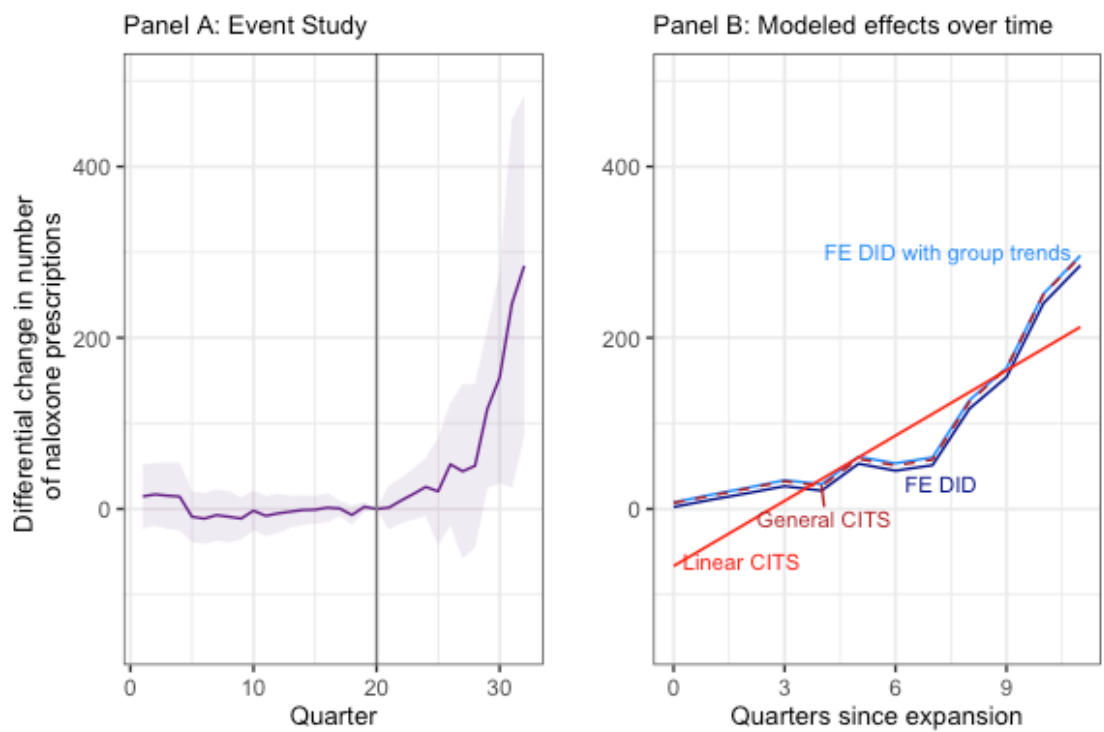

SOURCES/NOTES: **Sources** Authors' re-analysis of Medicaid covered naloxone prescriptions from Frank & Fry, 2019. **Notes** Panel A is an event study plot, where the adjusted differential in outcome between the treated group and comparison group relative to the time of intervention is estimated for each time period before and after Medicaid expansion. Panel B provides time-varying estimates for each month after Medicaid for FE DID, FE DID with group trends, general CITS, and linear CITS. Covariate adjustment is the same for each model presented in both Panels A and B and is exactly the same as the covariate adjustment used in Frank & Fry, 2019.

**Figure 4.** Comparison of Estimates Across Study Designs - Powell, et al., 2019

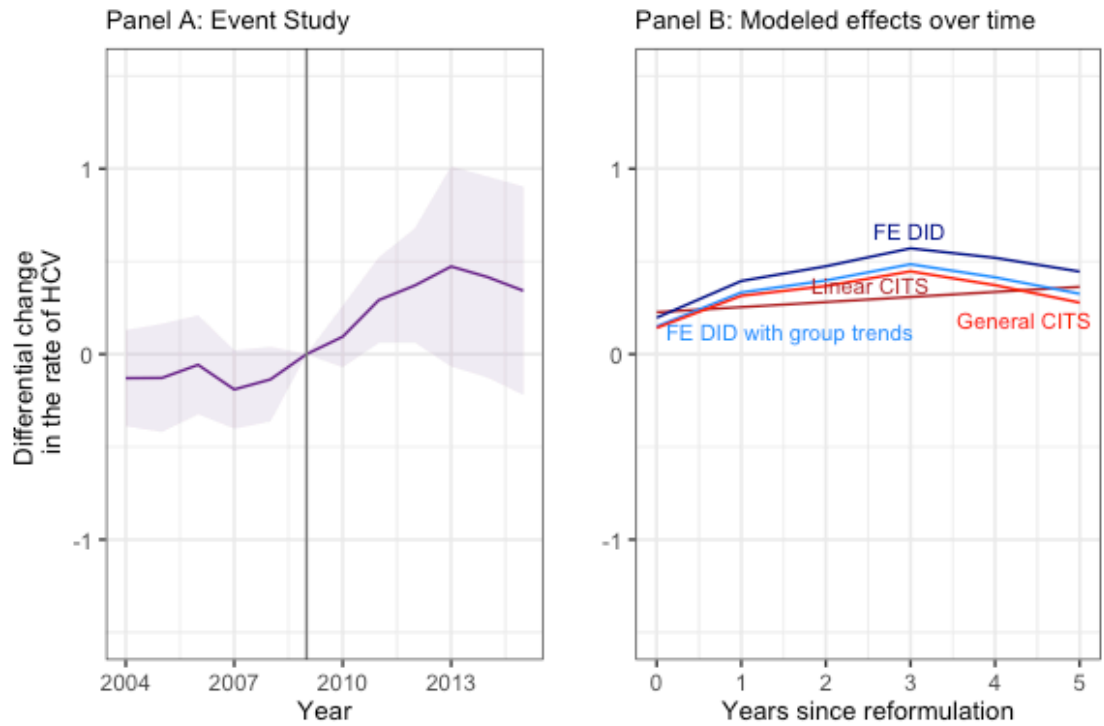

SOURCES/NOTES: **Sources** Authors' re-analysis of acute HCV incidence from Powell et al., 2019. **Notes** Panel A is an event study plot, where the adjusted differential in outcome between the treated group and comparison group relative to the time of intervention is estimated for each time period before and after Medicaid expansion. Panel B provides time-varying estimates for each month after Medicaid for FE DID, FE DID with group trends, general CITS, and linear CITS. Covariate adjustment is the same for each model presented in both Panels A and B and is exactly the same as the covariate adjustment used in Powell et al., 2019. Unlike Powell et al., 2019 who use a continuous treatment variable, we use a binary exposure variable with the cutoff for "high exposure" being the median OxyContin misuse rate in the pre-period.

21
Fry & Hatfield, 2020